\documentclass[12pt]{article}
\usepackage{amsfonts}
\usepackage{amssymb}
\usepackage{graphics,psboxit,amsmath}

\def\hybrid{\topmargin -20pt    \oddsidemargin 0pt
        \headheight 0pt \headsep 0pt
        \textwidth 6.35in       
        \textheight 9.25in       
        \marginparwidth .875in
        \parskip 5pt plus 1pt   \jot = 1.5ex}

\hybrid

\def\baselinestretch{1.2}

\catcode`\@=11

\def\marginnote#1{}
\newcount\hour
\newcount\minute
\newtoks\amorpm
\hour=\time\divide\hour by60
\minute=\time{\multiply\hour by60 \global\advance\minute by-\hour}
\edef\standardtime{{\ifnum\hour<12 \global\amorpm={am}%
        \else\global\amorpm={pm}\advance\hour by-12 \fi
        \ifnum\hour=0 \hour=12 \fi
        \number\hour:\ifnum\minute<10 0\fi\number\minute\the\amorpm}}
\edef\militarytime{\number\hour:\ifnum\minute<10 0\fi\number\minute}

\def\draftlabel#1{{\@bsphack\if@filesw {\let\thepage\relax
   \xdef\@gtempa{\write\@auxout{\string
      \newlabel{#1}{{\@currentlabel}{\thepage}}}}}\@gtempa
   \if@nobreak \ifvmode\nobreak\fi\fi\fi\@esphack}
        \gdef\@eqnlabel{#1}}
\def\@eqnlabel{}
\def\@vacuum{}
\def\draftmarginnote#1{\marginpar{\raggedright\scriptsize\tt#1}}

\def\draft{\oddsidemargin -.5truein
        \def\@oddfoot{\sl preliminary draft \hfil
        \rm\thepage\hfil\sl\today\quad\militarytime}
        \let\@evenfoot\@oddfoot \overfullrule 3pt
        \let\label=\draftlabel
        \let\marginnote=\draftmarginnote
   \def\@eqnnum{(\theequation)\rlap{\kern\marginparsep\tt\@eqnlabel}%
\global\let\@eqnlabel\@vacuum}  }

\def\preprint{\twocolumn\sloppy\flushbottom\parindent 2em
        \leftmargini 2em\leftmarginv .5em\leftmarginvi .5em
        \oddsidemargin -.5in    \evensidemargin -.5in
        \columnsep .4in \footheight 0pt
        \textwidth 10.in        \topmargin  -.4in
        \headheight 12pt \topskip .4in
        \textheight 6.9in \footskip 0pt
        \def\@oddhead{\thepage\hfil\addtocounter{page}{1}\thepage}
        \let\@evenhead\@oddhead \def\@oddfoot{} \def\@evenfoot{} }

\def\numberbysection{\@addtoreset{equation}{section}
        \def\theequation{\thesection.\arabic{equation}}}

\def\underline#1{\relax\ifmmode\@@underline#1\else
        $\@@underline{\hbox{#1}}$\relax\fi}

\def\titlepage{\@restonecolfalse\if@twocolumn\@restonecoltrue\onecolumn
     \else \newpage \fi \thispagestyle{empty}\c@page\z@
        \def\thefootnote{\fnsymbol{footnote}} }

\def\endtitlepage{\if@restonecol\twocolumn \else \newpage \fi
        \def\thefootnote{\arabic{footnote}}
        \setcounter{footnote}{0}}  

\catcode`@=12
\relax

\def\figcap{\section*{Figure Captions\markboth
        {FIGURECAPTIONS}{FIGURECAPTIONS}}\list
        {Figure \arabic{enumi}:\hfill}{\settowidth\labelwidth{Figure
999:}
        \leftmargin\labelwidth
        \advance\leftmargin\labelsep\usecounter{enumi}}}
 \relax
\def\tablecap{\section*{Table Captions\markboth
        {TABLECAPTIONS}{TABLECAPTIONS}}\list
        {Table \arabic{enumi}:\hfill}{\settowidth\labelwidth{Table
999:}
        \leftmargin\labelwidth
        \advance\leftmargin\labelsep\usecounter{enumi}}}
 \relax
\def\reflist{\section*{References\markboth
        {REFLIST}{REFLIST}}\list
        {[\arabic{enumi}]\hfill}{\settowidth\labelwidth{[999]}
        \leftmargin\labelwidth
        \advance\leftmargin\labelsep\usecounter{enumi}}}
 \relax
\makeatletter
\newcounter{pubctr}
\def\publist{\@ifnextchar[{\@publist}{\@@publist}}
\def\@publist[#1]{\list
        {[\arabic{pubctr}]\hfill}{\settowidth\labelwidth{[999]}
        \leftmargin\labelwidth
        \advance\leftmargin\labelsep
        \@nmbrlisttrue\def\@listctr{pubctr}
        \setcounter{pubctr}{#1}\addtocounter{pubctr}{-1}}}
\def\@@publist{\list
        {[\arabic{pubctr}]\hfill}{\settowidth\labelwidth{[999]}
        \leftmargin\labelwidth
        \advance\leftmargin\labelsep
        \@nmbrlisttrue\def\@listctr{pubctr}}}
 \relax
\makeatother
\newskip\humongous \humongous=0pt plus 1000pt minus 1000pt

\newif\ifdtup

\relax

\def\be{\begin{equation}}
\def\ee{\end{equation}}
\def\ba{\begin{eqnarray}}
\def\ea{\end{eqnarray}}

\def\no{\noindent}

\def\IR{\relax{\rm I\kern-.18em R}}
\def\II{\relax{\rm 1\kern-.35em1}}

\renewcommand{\theequation}{\thesection.\arabic{equation}}
\csname @addtoreset\endcsname{equation}{section}

\def\IR{\relax{\rm I\kern-.18em R}}
\def\inv{^{\raise.15ex\hbox{${\scriptscriptstyle -}$}\kern-.05em 1}}

\begin{document}

\begin{titlepage}
\begin{center}

\hfill CERN-PH-TH/2008-226\\
\vskip -.1 cm
\hfill IFT-UAM/CSIC-08-79\\
\vskip -.1 cm
\hfill arXiv:yymm.nnnn\\

\vskip .5in

{\LARGE Quantum Information and Gravity Cutoff in Theories with  Species }
\vskip 0.4in

{\bf Gia Dvali$^{1,2}$ and Cesar Gomez$^{3}$}

\vskip 0.1in

${}^1\!$
Theory Division, CERN\\
CH-1211 Geneva 23, Switzerland\\
${}^2\!$ Center for Cosmology and Particle Physics,\\
Department of Physics,\\
New York University, New York,  NY 10003, USA\\


${}^3\!$
Instituto de F\'{\i}sica Te\'orica UAM-CSIC, C-XVI \\
Universidad Aut\'onoma de Madrid,
Cantoblanco, 28049 Madrid, Spain\\
{\footnotesize{\tt georgi.dvali@cern.ch , cesar.gomez@uam.es}}

\end{center}

\vskip .4in

\centerline{\bf Abstract}
\vskip .1in
\no

\noindent
We show that lowering of the gravitational cutoff relative to the Planck mass, imposed by black hole physics in theories  with $N$ species, has an independent justification from quantum information theory. First, this scale marks the 
limiting capacity of any information processor.  Secondly, by taking into the account the limitations of  the quantum information storage in any system with species,  the bound on the gravity cutoff becomes equivalent to the holographic bound, and this equivalence automatically implies the equality of 
entanglement and Bekenstein-Hawking entropies.  Next, the same bound follows from 
quantum cloning theorem.  Finally, we point out that by identifying the UV and IR  threshold scales of the black hole quasi-classicality in four-dimensional field  and high-dimensional gravity theories, 
the bound translates as the correspondence between the two theories.  In case when the high-dimensional background is AdS,  this reproduces the well-known AdS/CFT relation, but also suggests
a generalization of the correspondence beyond AdS spaces.   In particular, it reproduces a recently suggested duality  between a four-dimensional CFT and a flat five dimensional theory, in which  gravity crosses over from four to five dimensional regime in far infrared.

\vskip .4in
\noindent

\end{titlepage}
\vfill
\eject

\def\baselinestretch{1.2}


\baselineskip 20pt


\section{Introduction}

\no

  In the presence of $N$ elementary particle species in an effective  quantum field theory, the consistency of the  large-distance black hole physics imposes the following bound on the gravitational cutoff of the theory \cite{bound}
   \be\label{bound}
 \Lambda \, = \, {M_{Pl} \over \sqrt{N}} \,  .
\ee 
Thus, in the presence of $N$ particle species, the fundamental length is no longer $l_P \equiv M_{Pl}^{-1}$, but rather
\begin{equation}
\label{ln}
l_{N} \, \equiv\,  \sqrt{N}/M_{Pl}\, .
\end{equation}
   This fact can be seen by a number of different arguments, some of which will be briefly reproduced below.  The clearest indication of 
(\ref{bound}) is  perhaps the fact that 
in the presence of $N$ species, the black holes of size $\ll l_N$ cannot be Einsteinian, since otherwise they would half-evaporate within the time shorter than their size, which is impossible. 
Another indication is the consistency between the entanglement and Bekenstein-Hawking entropies
\cite{entropy} (see below).     
The black hole arguments are qualitatively  supported by the perturbation theory intuition \cite{perturbative, gabriele, redi}, since species do contribute into the  renormalization of the 
gravitational constant. However, the perturbation theory alone cannot lead to any rigorous bound, since 
it is not  protected against the cancellations among the tree-level and the loop contributions, as well as cancellations among the species of different spin, or simply against the re-summation of the perturbative  series.  Below, we shall rely solely on  fully non-perturbative  treatment. 
   
 In the present paper we wish to shed a different light  on the black hole  bound on the gravity cutoff (\ref{bound}), from the point of view of the quantum information theory.  The species label represents a particular form of information, and absorbtion and emission of the species by a black hole is  a particular form of the information processing.   At the same time,  the  species back-react on the processing capacity of   
a black hole, and this back reaction is sensitive to their number.  
   
 Thus, it is natural to ask whether there is an underlying  connection between the bound (\ref{bound}) and the information   storage and processing in the presence of black holes.  It is the purpose of the present note to establish the above connection.  The summary of our findings is as follows. 
    
     First,  we shall show that the distance  
  $l_N$ sets the shortest space-time volume within which the identification of species is in principle possible.   Thus,  it  imposes the  limit on the decoding capacity of any detector that can read information stored in the species. This provides an independent evidence for (\ref{bound}).   
  
 Secondly,  we show that by consistently defining the cutoff as the minimal length (time) scale in which one can  store and process the maximal number of independent bits of the information encoded in the species labels,  the holographic bound automatically translates as (\ref{bound}), and explains 
equality between the entanglement and Bekenstein-Hawking entropies.   Alternatively, by 
requiring equality between the two entropies, the holographic bound and (\ref{bound}) become 
equivalent.  
   
    Next, we focus on the quantum information processing and  
will show, that the bound (\ref{bound}) follows from the quantum cloning theorem \cite{Susskind}. 

 Finally, we shall discuss the connection between (\ref{bound}) and  CFT/gravity  correspondence. 
We show,  that  by requiring the equality of UV and IR scales beyond which the black holes go out of the
respective Einsteinian regimes in four-dimensional field theory with $N$-species, on one side,  and 
higher-dimensional pure-gravity theory on another,  a certain well-defined correspondence between the two theories follows.  In a particular case,  when the high-dimensional background is AdS, this reproduces the well known\cite{Maldacena}  AdS/CFT relation.  However, the formulation in terms of the 
black hole regimes includes no explicit reference to the curvature of the high-dimensional background  
but only to the crossover scales between the different  gravitational regimes. Therefore, such a formulation suggests a way of generalizing the correspondence beyond AdS spaces.  As an evidence, we show that this formulation correctly reproduces the recently suggested duality \cite{barvinski}  between four-dimensional 
CFT with $N$ species, and five-dimensional gravity theory of type \cite{dgp}, in which gravity changes the regime from four to five dimensional behavior at the infrared crossover scale $r_c \, = \, \sqrt{N}/M_P$.


\section{Black Hole Evaporation Argument}

 We shall first  briefly reproduce the black hole evaporation argument\cite{bound,redi}, leading to (\ref{bound}). 
We can prove that the scale (\ref{bound}) marks the gravity cutoff, by showing that any black hole 
of the size $ R \, \lesssim \, l_N$,  cannot represent a quasi-classical Einsteinian black hole.  Notice,  that the necessary condition for quasi-classicality is,  that the rate of the temperature-change is less than the temperature squared, or equivalently, the evaporation time has to be longer than the black hole size $R$. 
 This condition is impossible to satisfy for the quasi-classical  Schwarszcvhild black holes of size  $R \, \lesssim  \, l_N$.     
 Indeed, if such a black hole could be an approximately Schwarszchildian, 
 its half-evaporation time would go as,   
\be\label{four}
\tau \, \sim  \, R^3M_{Pl}^2/N \,  \lesssim \, R\, , 
\ee
where in the very last inequality we have used the definition of $l_N$ and the fact that it exceeds $R$. 
Thus, the quasi-classicality condition is inevitably violated. Hence no quasi-classical Schwarszchild black hole can exist  below the size $l_N$, which proves that the latter  length is the gravity cutoff.   

 Notice that a potential instability of the species cannot provide a loophole from the above reasoning, 
 since in the evaporation of an Einsteinian black hole of size $\Lambda^{-1}$, all the species 
 with the mass $M\, < \, \Lambda$ contribute,  irrespectively of their exact lifetime. 
 To see this, consider such a black hole which is at rest in a lab reference frame, and assume that the 
 species have a  mass $M$ and a decay width $\Gamma$.  Since by assumption the black hole is Einsteinian,  the evaporation is thermal, and the species are emitted by a typical energy and  momenta 
$\sim \Lambda$. Thus, in the lab frame, the lifetime of an emitted particle is prolonged  by the usual Lorenz factor, 
\be
\tau_{lab} \, = \,  \Gamma^{-1} {\Lambda \over  M} \, . 
\ee
Since,  for any elementary particle  by default $\Gamma \,  <  \, M$ (or else the state is 
not a particle, but  rather a broad resonance),  this time scale is longer than the black hole lifetime, computed in Einsteinian theory.   Hence,  all the particles count, and the bound is insensitive to the 
particle instability.   

 We shall now proceed to investigate the fundamental meaning of the scale (\ref{bound}) from the quantum information point of view.

\section{The Fastest Processors} 

  We first wish to show that the scale $l_{N}$ sets the 
shortest possible time and length scale, over which the proper identification of the species is possible. Consider a theory with $N$ elementary particle species,  $\Phi_j$,  labeled by an index $ j \, =\, 1,2, ...N$.
  For the beginning we shall assume that the species are stable. One can then store an information in these species. For example, the simplest form of the information is the number of quanta of a given species. If species are stable, and the associated species number is exactly conserved (modulo some periodicity), the information can be stored arbitrarily long in form of a set of these numbers. 
For example, without loss of generality, let the exactly conserved quantum number for $j$-th species have periodicity $n_j$. This means,  that there are $N$ exact discrete symmetries $Z_{n_j}$, 
under which the corresponding species transform as, 
\be\label{species}
\Phi_j \, \rightarrow \, {\rm e}^{i{2\pi \over n_j}}\Phi_j \,.
\ee 
In such a case an {\it everlasting} information can be stored in the form of the set of $N$ numbers $n_1,\, n_2\, ...\, n_j\, ...\, n_N$. 
In the absence of gravity, there are no limits to either  $n_j$ or $N$. Both, the individual numbers and their multiplicity can be arbitrarily large. In addition, if 
species are bosons, the information can be stored in an arbitrarily small volume (in case of the fermions, the obvious limitations apply due to Pauli's exclusion principle). However, what is more important for our present analysis, in an imaginary world without gravity this information can also be decoded arbitrarily fast.  Indeed, consider a simplest form of such information, carried by a single quantum of 
$\Phi_j$. This information is encoded in the number set $n_j\, = \, \delta_{j1}$.  In order to decode this information, all we have to do is to recognize the index $j$. The particle detector that performs this task must contain a presorted sample of particles (and antiparticles) from all the available $N$ species, and simply detect with which of these particles the probe 
$\Phi_j$ interacts in some recognizable way (e.g, annihilates). 

 In the absence of gravity, the size of the pixel that stores the $N$ sample particles can be arbitrarily small, and the detection process can correspondingly be arbitrarily fast.  
  However, gravity puts an inevitable limit to the smallness of the pixel, and thus, to 
the decoding capacity of any detector. Indeed, the typical momentum of a sample particle localized within a pixel of size $l_{pix}$, is $P \sim l_{pix}^{-1}$. In Einsteinian gravity 
the pixel with $N$ overlapping such particles will collapse into a black hole when the gravitational radius of the pixel exceeds its size. This will happen for 
\be\label{radius}
l_{pix}^2 \, < \, {N \over M_{Pl}} \, ,
\ee 
which coincides with the  bound  (\ref{bound}).
Thus, the scale $l_N$ sets the ultimate lower bound on the size of a pixel, and thus, on 
a decoding capacity of the detector.  
 
 Notice that making a multi-layer pixel detector, in which the sample species are spread over a distance 
 larger than the gravitational radius of the pixel,  does not offer the way out, since such a detector will inevitably increase the detection time beyond $l_N$. 

  We can now generalize this reasoning to the unstable species. Qualitatively it is clear that 
instability of the species can only affect the long term capacity of the information storage, 
because of the finite lifetime. But the short-term storage should not  be affected. 
This is indeed the case, as we shall now demonstrate.  Let the masses and the decay widths of the species be $M$ and    $\Gamma$ respectively (for simplicity we assume this parameters to be 
species-independent, generalization to non-universal masses is straightforward).  Since species
are elementary particles, we must have $M \gg \Gamma$, otherwise the species are broad resonances, and  should not be counted. 

 Because species are unstable, the species number symmetries $Z_{n_j}$ can no longer be exact. 
Under the species number symmetry we mean the symmetries that  are ascribed only to the species
and not to their decay products, even if some other exact symmetry is carried by the latter. (For example, in neutron beta decay, the neutron number is violated, although the baryon number is conserved and is transfered to the proton). 
 However,  in the rest-frame of a particle, the species numbers  are still approximately conserved  on the time-scales  $t \ll \Gamma^{-1}$, and they can be used for the information storage on such scales.  
 In the example of beta decay, the neutron number is approximately conserved on the time scales  $t \ll  10^3$ sec, and we can reliably store information into the neutron number during the shorter time. 
For much later  times  the neutron number is no longer conserved, and cannot be used for the information storage.  Instead, the baryon number can be reliably used for the information storage on much longer time scales,  at least up to $10^{34}$y, the current experimental bound on the proton 
lifetime.   

 Now it is clear that the instability of the species cannot improve the decoding capacity of the detector, 
 since for the decoding to work the necessary condition is that the size of the pixel must be 
 $l_{pix} \, \ll \,  \Gamma^{-1}$, otherwise the detector itself will decay before any reading of information 
becomes possible.   Then, everything goes as for the stable species. 

 We see that the bound on the gravity cutoff (\ref{bound}), that is suggested by the black hole evaporation physics,  is also a bound on the decoding capacity of any detector for the system in which information is stored 
 in the number of species.   This fact suggests a non-trivial underlying connection between the 
way black holes  processes the information and the number of particle species.  We shall now turn to another aspect of this connection.

\section{Quantum Cloning and Gravity Cutoff}
\subsection{Gravity Cutoff and Entanglement}
Entanglement entropy for a system divided by a surface $\Sigma$ depends on the area of $\Sigma$ measured in units of the ultraviolet cutoff $\Lambda$ as well as on the number $N$ of different particle species as
\be\label{one}
S=N\Lambda^2 A(\sigma)
\ee 
If we identify the ultraviolet cutoff with the gravity scale $M_{Pl}$ the entanglement entropy becomes the universal entropy for a black hole with horizon $\Sigma$ multiplied by the number of species. Since a black hole horizon is a natural physical way to define the conditions of entanglement, the only way to reconcile the Bekenstein-Hawking entropy with the entanglement entropy (\ref{one}) is assuming a new gravitational cutoff depending on the number of species, namely
(\ref{bound}).   This result is supported by explicit examples in which the relation (\ref{bound}) as well as the value of both entropies are calculable  from the fundamental theory  \cite{entropy}.
The relation 
$\Lambda^{2}N=M_{Pl}^2$ has on the other hand a simple quantum information interpretation. In fact, if we take $\Lambda$ as $M_{Pl}$ and we interpret $S$ as the number of Boolean degrees of freedom \cite{tHooft}, then (\ref{one}) will imply that we can pack $N$ bits of information in one Planck unit cell in contradiction with the holographic bound. In order to solve this contradiction we need to introduce the scale (\ref{bound}).

We can arrive to the same conclusion by the following reasoning.  As discussed in Sec. 2,  we can consistently  define the cutoff length as the minimal space-time scale $\Lambda^{-1}$,  on which one 
can store and process the information encoded in $N$ distinct species.  Since,  in each species-number we can encode at least one bit of the information ($n_j \, = \, \delta_{j1}$), we can thus encode total of 
$N$ bits of such information in an area $\Lambda^{-2}$.  But, the holographic bound implies 
that  this maximal number should be equal to the number of the Planck area  pixels $M_P^{-2}$
per same surface.  That is,  $N \, = \, M_P^2/\Lambda^2$.  Thus,  the holographic principle, when 
confronted with the information storage capacity of  $N$-species theory, automatically implies (\ref{bound}) and consequently the equality of entanglement and Bekenstein-Hawking entropies.   
 
 We can reverse the connection in different ways.  In particular, requirement of  equality between the 
 Bekenstein-Hawking and the entanglement entropies as the starting point, implies equivalence between the holographic and (\ref{bound})  bounds.  
  A complementary important question, not addressed in this work,  it the underlying connection between $N$ and the entropy bounds\cite{private}.

\subsection{Gravity Cutoff and Quantum Cloning}
Lower bounds on black hole evaporation time can be independently established using quantum information theory, namely quantum cloning \cite{Susskind}. Using Rindler coordinates for the black hole horizon we can use the well known Alice-Bob experiment. Denoting 
$\omega$ the Rindler retrieval time, Bob will jump into the black hole with the information about Alice quantum state at 
\be
X^{+}=R e^{\omega}
\ee
Since the singularity is at $X^{+}X^{-}=R^{2}$ for $R$ the black hole Schwarszchild radius, quantum cloning will take place if Alice can send the information to Bob in a time interval of the order 
\be
R e^{-{\omega}}
\ee
By uncertainty principle this requires an energy of the order of $R^{-1}e^{\omega}$ that should be smaller that the black hole mass. Therefore quantum cloning will take place if the retrieval time $\omega$ is smaller than $\log(R)$. Since uncertainty principle implies the impossibility of quantum cloning we get the following lower bound on black hole retrieval time
\be\label{five}
\omega > \log(R)
\ee
This bound is clearly satisfied if according to (\ref{four}) we consider that the retrieval time is of the order $R^{2}$. However the situation changes once we consider that we have $N$ different species. In this case the retrieval time will be reduced to $\frac{R^{2}}{N}$ and the cloning bound will become
\be
\frac{R^{2}}{N} > \log(R).
\ee
Restoring the powers of $M_{Pl}$, this is obviously violated if the gravity cutoff is above the scale ${M_{Pl} \over \sqrt{N}}$. 
Indeed, let us assume that the cutoff can be much higher,   $\Lambda^{2} \, \gg \,  M_{pl}^{2}/N$, and apply the above Bob and Alice experiment to a black hole of an intermediate size  
$\Lambda^{-1} \,  \ll \,  R  \, \ll  \,  \sqrt{N}/M_{Pl}$. Since by assumption  the size of  such a 
black hole is bigger than the cutoff length  $\Lambda^{-1}$, the black hole in question  must be 
nearly Einsteinian, and the usual thermal evaporation rates apply.  
Such a black hole then would half-evaporate within the time (\ref{four}). 
But this time is less than the times that Alice and Bob need to reach  the singularity.  Thus,  Bob would have enough time to retrieve the information from the Hawking radiation, then fall into a black hole and get a message from Alice,  thus creating a problem with cloning.  The only way out is to  assume 
that for $R \ll l_N$ black holes cannot be Einsteinian, which implies the bound (\ref{bound}).  

 To summarize,  the essence of our argument is the following.  In Alice and Bob experiment the time scales involved are: {\it (a)} The time that  is available for Alice to send a message to Bob before she  reaches  the singularity (call it  $t_A$);   {\it (b)} The time is takes Bob to retrieve the information from the Hawking radiation (call it $t_{H}$); and finally, {\it (c)}  The minimal time  for Bob to fall into the black hole and reach the singularity  (call it $t_B$). 

 In Schwarszchild coordinates, 
 for a quasi-classical Einsteinian  black hole of size $R$,  the times  $t_A$ and $t_B$ are of order 
 $R$ at the starts of the falls correspondingly.   The  retrieval time on the other hand is of order the half-evaporation time (\ref{four}), $t_H\, \tau \,  \sim \, R^3M^2_{Pl}/N$. The problem with cloning appears because for an Einsteinian  black hole of size 
$R \ll l_N \equiv \sqrt{N}/M_{Pl}$,  we have $t_A\sim t_B \, \gg \, t_H$.
 For a black hole of this size,  the available `lifetimes'  for both Alice and Bob are of the same order and of order $R$, whereas the information retrieval  time for Bob is much shorter. This is the crucial role of large $N$. In other words, because of accelerated 
half-evaporation time due to emission of too many species,  Bob can retrieve the information very fast and 
follow Alice with just a minor delay. Alice, therefore,  has enough time to send him a copy of the information  using quanta  as soft as of the  wavelength $\sim t_A$ and energy 
$1/t_A$. Because  $t_A \sim R$,  for an Einsteinian black hole  the energy involved would {\it by default} be smaller than the  black hole mass  $M_{BH} \,  \sim \, RM_{Pl}^2 \, \gg \, R^{-1}$.   Thus, Bob would have all the conditions for creating the problem with cloning. 

 The only way out is to accept that the black holes smaller than $l_N$  are no longer Einsteinian, which implies  the gravity cutoff given by (\ref{bound}).   
 In summary we conclude that the gravity cutoff in presence of different species is determined by quantum cloning theorem.


\subsection{Scrambling time}
In \cite{HaydenPreskill} a new time scale for black hole emission of information was introduced. This is the scrambling time that very likely can be identified \cite{susskind2} with the time required for the diffusion 
of a bit of information over the entire horizon. In Schwarzschild coordinates this time is determined by
\be
e^{-\frac{t}{D}} = R
\ee
leading to $t \sim D \log(R)$. For the membrane model of the stretched horizon and based on dissipative properties of classical hair as electric charge we get for the diffusion coefficient $D=\frac{1}{4\pi T}= R$ implying a Rindler scrambling time $\omega$ of order $\log (R)$.

The main result of \cite{HaydenPreskill} is that the retrieval time for Bob to get Alice's information could be as small as the 
scrambling time, saturating the bound $\omega > \log (R)$ impossed by the cloning theorem. If we consider the existence of $N$ different species and we decide to keep the gravity scale as $M_{Pl}$ we will arrive to the paradoxical situation that , for large enough $N$, the retrieval time $\frac{R^{2}}{N}$ could be smaller than the scrambling time. A priori a potential way out of this puzzle, without introducing a new gravity scale, would be to change the diffusion coefficient $D$ to $\frac{D}{N}$. However if we assume the membrane model for the stretched horizon we can write $D$ in an hydrodynamical approximation as
\be
D \sim \frac{\eta}{\epsilon +p} = \frac{\eta}{ST}
\ee
for $\eta$ the viscosity and therefore a rescaling of $D$ depending on the number of species will modify the viscosity-entropy relation \cite{son}
$\frac{\eta}{S}=\frac{1}{4\pi}$ based on AdS-CFT correspondence. Thus we conclude that the scrambling time is independent on the number of species which together with the cloning theorem leads us again to identify the gravity scale with $\frac{M_{Pl}}{\sqrt{N}}$.

\section{AdS/CFT and the Gravity Bound}
The AdS/CFT correspondence \cite{Maldacena} is build up on the relation\footnote{Notice that in this
section, the number of species is $N^2$ rather than $N$. So whenever used  below,  in (\ref{bound})  we have to replace $N \, \rightarrow \,  N^2$.} 
\be\label{one1}
R= l_{s}(g_{s}N)^\frac{1}{4}
\ee
between the curvature radius of bulk AdS gravitational background and the gauge t«Hooft coupling. Strictely speaking relation (\ref{one1}) defines the characteristic gravity scale of a set of $N$ black 3-branes. Recall that the corresponding metric is $ds^{2}=H^{-\frac{1}{2}}dx.dx + H^{\frac{1}{2}}dydy$ with $H=1+(\frac{R}{r})^{4}$ with $R$ given by (\ref{one1}) once the D-brane tension is determined.The holographic meaning of (\ref{one1}) is based on the IR/UV correspondence between bulk gravity and boundary gauge theory \cite{SusskindWitten}. In fact if we consider the four dimensional gauge theory with $N^{2}$ species defined on $S^{3} \otimes R$ with $S^{3}$ of radius equal one and we introduce an UV cutoff $\Lambda_{UV}$, the total number of degrees of freedom of the theory is given by
\be
N^{2} \Lambda_{UV}^{3}
\ee
The holographic principle implies that the total number of degrees of freedom should be equal to the area, measured in five-dimensional Planck units, of a regularized boundary in AdS for some IR cutoff $\Lambda_{IR}$
\be\label{two2}
N^{2} \Lambda_{UV}^{3} = R^{3} \Lambda_{IR}^{3} M_{Pl}^{3}
\ee
where $R^{3} \Lambda_{IR}^{3}$ is the area of the sphere at $r=1-\frac{1}{\Lambda_{IR}}$ for the AdS metric
\be
ds^{2}=R^{2} ( \frac{4dx_{i}dx_{i}}{(1-r^{2})^{2}} - dt^{2} \frac{1+r^{2}}{1-r^{2}})
\ee
Since the IR cutoff in the bulk induces an UV cutoff on the boundary theory we can identify $\Lambda_{UV}=\Lambda_{IR}$ in (\ref{two2}) which leads to (\ref{one1}) once we define the five-dimensional Planck mass by standard KK compactification on $S^{5}$ with radius equal $R$,
$M_{Pl}^{3}=R^{5}\frac{1}{l_{s}^{8}g_{s}^{2}}$.
Thus (\ref{one1}) becomes equivalent to
\be\label{three3}
N^{2} \Lambda^{3} = M_{Pl}^{3}
\ee
for $\Lambda=\frac{1}{R}$. It is amusing to observe now that if we
define the five-dimensional Planck mass as
$M_{Pl} ^{3}=M_{Pl4}^{2} \frac{1}{R}$
the correspondence (\ref{three3}) becomes exactly the gravity bound relation
\be\label{four4}
N^{2} \Lambda^{2}= M_{Pl4}^{2}
\ee
for a four-dimensional theory with $N^{2}$ species with the AdS radius playing the role of the gravity scale $l_{N}$.
Reciprocally we can start with the gravity bound (\ref{four4}) for a 
four-dimensional quantum field theory with $N^{2}$ species. If now we define $M_{Pl4}$ as
$M_{Pl4}^{2}= M_{Pl10}^{8} R^{6}$
for a generic Kaluza Klein scale $R$ and we write, in full generality, the gravity cutoff $\Lambda$ as $a\frac{1}{R}$ for some arbitrary constant $a$ the gravity bound leads to the relation
\be
R=a l_{s}(g_{s}N)^{\frac{1}{4}}
\ee
that for $a=1$ is the correspondence (\ref{one1}). In other words if for a four dimensional theory with $N^{2}$ species we compactify from ten to four dimensions (i.e we assume a string theory completion) with KK scale $R$ then the correspondence (\ref{one1}) becomes equivalent to
\be\label{lNR}
l_{N^2} = R, 
\ee
where,  $l_{N^2} \equiv M_P/N$, is the gravity bound scale for $N^2$ species. In summary we observe that the typical gravity scale $R$ for $N$ D-branes as given by (\ref{one1}) has the information theoretical meaning of being the shortest length scale $l_{N^2}$ over which identification of the different species attached to the D-branes is possible.

The above result can be formulated in the black hole language. 
Namely,  applying the bound (\ref{bound}) to four-dimensional field theory 
with $N^2$ species, we  can obtain AdS/CFT relation 
 by requiring the equality of UV and IR scales, beyond which the black holes of the two theories go out of the corresponding  Einsteinian regimes. 
  Indeed,   on the 4D field theory side with $N^2$ species, this is the scale 
 $\Lambda_{UV} = l_{N^2}^{-1}$, due to the black hole evaporation argument, reproduced in section 2.   Black holes smaller than $l_{N^2}$ evaporate too quickly and cannot 
 be quasi-classical.  
  On the side of  the 10D theory, both the compactification radii as well as the AdS curvature set the obvious IR scale 
 $\Lambda_{IR}  = R^{-1}$, below which no classical black holes exist \cite{cft}. 
  Equating the two scales and using the usual geometric relation between the four and ten--dimensional Planck masses and the string scale, we recover the well known AdS/CFT relation. 
  
  The  above formulation in the black hole language  carries no explicit reference to the background  
curvature of the high-dimensional theory, but only to the  UV  and IR crossover scales at which 
the change of the gravitational regime happens.  This suggests a way in which  
the field theory/gravity  correspondence may be extended beyond the AdS framework.  An interesting evidence of such an extension is a recently suggested duality \cite{barvinski} between the four-dimensional  CFT  with $N^2$ species and a flat five dimensional  
gravity theory of type \cite{dgp} with four-dimensional boundary Einstein term on the 3-brane. 
In the latter theory gravity changes from the four  to five dimensional regime 
at an IR crossover distance $r_c$.  It was observed in \cite{barvinski} that 
the scale factor evolutions {\it exactly} match in the two theories, subject to the identification,
\begin{equation}
     {N^2 \over M_P^2} \, = \,  r_c^2 \,.   
 \label{cftdgp}
 \end{equation}
Taking into the account the fact that $r_c^{-1} \, \equiv \, \Lambda_{IR}$ is the infrared scale 
at which gravity  (and correspondingly the black holes) changes the regime, and requiring UV-IR connection between the two theories, we realize that (\ref{cftdgp}) translates as the 
bound (\ref{bound})!

 Finally it is interesting to compare the gravity bound relation with the string-black hole 
 correspondence \cite{HorowitzPolchinski, DamourVeneziano}. If we denote $n$ the string multiplicity level then the mass of the corresponding string state is $\frac{n}{l_{s}^{2}}$. The string-black hole correspondence appears when we identify this mass with the mass for a black hole of radius of the order of the string lenght. This identification leads to the relation
\be
l_{s}=\frac{n^{\frac{1}{4}}}{M_{Pl}}
\ee
which implies that $l_{s}$ is just $l_{N}$ if we formally identify the number of species $N$ with the string multiplicity level $\sqrt{n}$. In the  same way that for the black hole string transition the critical value of the string coupling is at $g_{c}=n^{-\frac{1}{4}}$ we get from the gravity bound (\ref{bound}) a critical value for the string coupling 
$g_{c}=N^{-\frac{1}{2}}$ for $N$ species at which $l_{N} =l_{s}$.


\vspace{5mm}
\centerline{\bf Acknowledgments}
We would like to thank Luis Alvarez Gaume, Jose Barbon, Andrei Barvinski, Eliezer Rabinovici, Gabriele Veneziano for useful discussions.
The work of G.D is supported in part
by David and Lucile  Packard Foundation Fellowship for  Science
and Engineering, NSF grant PHY-0245068 and by the European Commission under 
the  "MassTeV"  ERC Advanced Grant 226371. The work of C.G. has been partially supported by the Spanish
DGI contract FPA2003-02877 and the CAM grant HEPHACOS
P-ESP-00346. This work was initiated during the TH Institute "Black Holes: A landscape of theoretical problems" . One of us (C.G) would like to thank CERN-Th for hospitality and partial support.





\begin{thebibliography}{99}
\bibitem{cft}
The saturation of the bound (\ref{bound}) on the CFT side due to  AdS/CFT correspondence between 
5D and 4D theories was  already noted in the first paper in \cite{redi}.  The difference is in the interpretation of the gravitational regime above $\Lambda_{UV}$.  There gravity was assumed to remain weakly coupled and classical above $\Lambda_{UV}$.  Our current interpretation of this scale 
is different.  Because we employ UV-IR connection,  above $\Lambda_{UV}$ no quasi-classical black holes may exist.
\end{thebibliography}
\end{document}